%% file: Main.tex
\documentclass[sigconf]{acmart}

\AtBeginDocument{%
  }

\setcopyright{acmlicensed}
\copyrightyear{2026}
\acmYear{2026}
\acmDOI{XXXXXXX.XXXXXXX}

\acmConference[CHI'26]{2026 CHI Conference on Human Factors in Computing Systems}{April 13--17,
  2026}{Barcelona, Spain}
\acmISBN{978-1-4503-XXXX-X/18/06}

\begin{document}

\title{Channelling, Coordinating, Collaborating:  A Three-Layer Framework for Disability-Centered Human-Agent Collaboration}

\author{Lan Xiao}
\affiliation{%
  \institution{Global Disability Innovation Hub}
  \institution{UCL Interaction Centre}
  \institution{University College London}
  \city{London}
  \country{United Kingdom}
}
\email{l.xiao.22@ucl.ac.uk}
\orcid{0009-0008-3468-7919}

\author{Catherine Holloway}
\orcid{0000-0001-7843-232X}
\affiliation{%
    \institution{Global Disability Innovation Hub}
     \institution{UCL Interaction Centre}
  \institution{University College London}
  \city{London}
  \country{United Kingdom}
}
\email{c.holloway@ucl.ac.uk}

\renewcommand{\shortauthors}{Xiao et al.}
\begin{teaserfigure}
  \includegraphics[width=\textwidth]{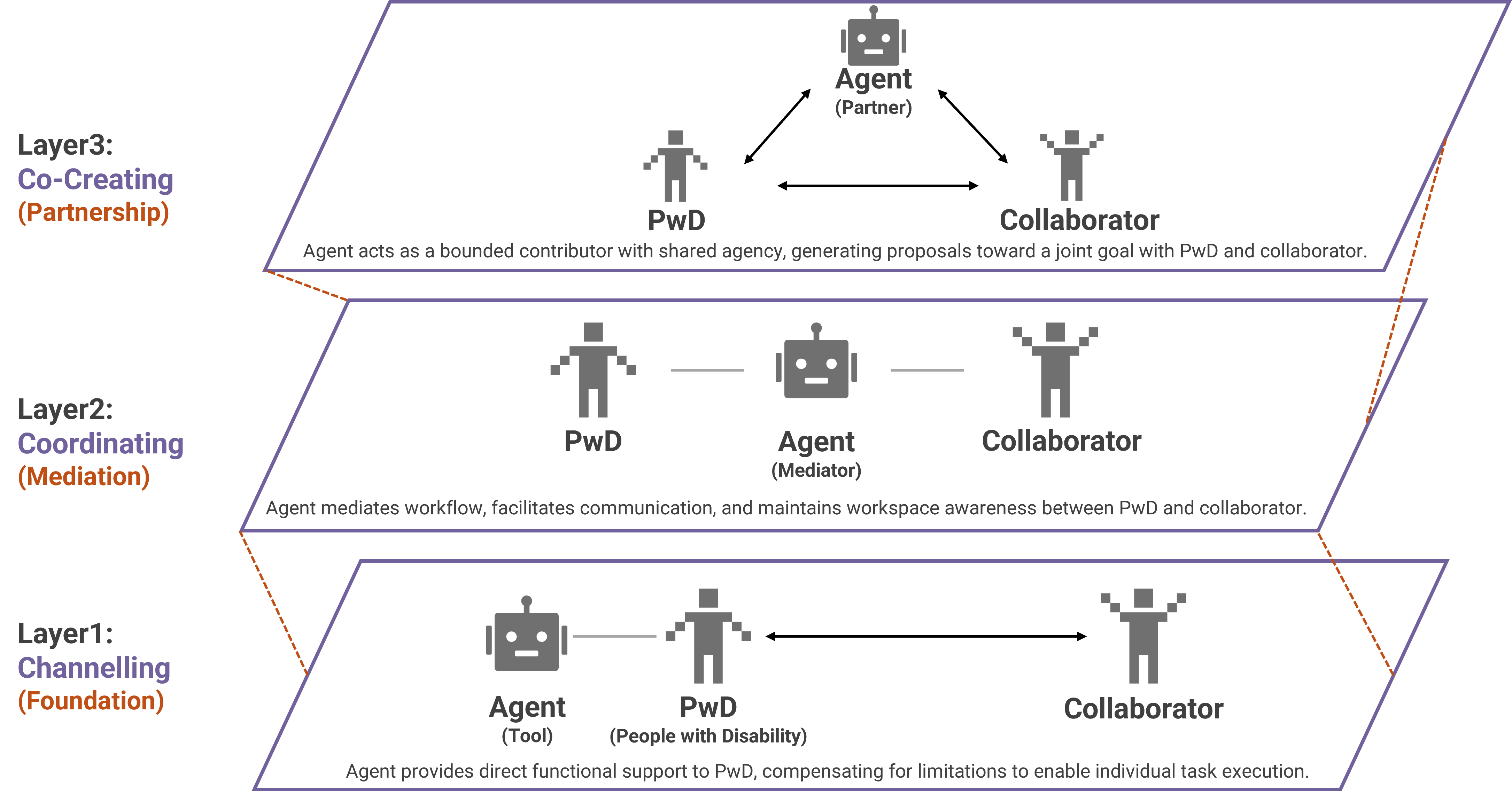}
  \caption{Disability-Centered Human-Agent Collaboration Framework}
  \Description{A three-layered framework diagram titled "Disability-Centered Human-Agent Collaboration," illustrating progressively complex interactions between a Person with Disability (PwD), an artificial intelligence Agent, and a human Collaborator, moving from foundational access to a co-creative partnership.}
  \label{fig:teaser}
\end{teaserfigure}
\begin{abstract}
AI accessibility tools have mostly been designed for individual use, helping one person overcome a specific functional barrier. But for many people with disabilities, complex tasks are accomplished through collaboration with others who bring complementary abilities, not solitary effort. We propose a three-layer framework, Channelling, Coordinating, and Co-Creating, that rethinks AI's role in ability-diverse collaboration: establishing shared informational ground across abilities, mediating workflows between collaborators with different abilities, and contributing as a bounded partner toward shared goals. Grounded in the Ability-Diverse Collaboration framework, grounding theory, and Carlile's 3T framework, it extends the ``agents as remote collaborators'' vision by centring the collaborative, interdependent ways people with disabilities already work.
\end{abstract}




\maketitle

\input{Chapters/1_Introduction}
\input{Chapters/2_Related_Work}
\input{Chapters/3_Background}
\input{Chapters/4_Method}
\input{Chapters/6_Discussion}
\input{Chapters/7_Conclusion}

\bibliographystyle{ACM-Reference-Format}
\bibliography{references}

\appendix

\end{document}

%% file: Chapters/1_Introduction.tex
\section{Introduction}
People with disabilities are increasingly using LLM-based tools across their everyday workflows, from information access \cite{adnin_i_2024} and day-to-day decision support \cite{tang_everyday_2025} to end-to-end content creation \cite{xiao_understanding_2025}. However, AI accessibility tools and much of the surrounding HCI research remain focused on individual assistance: compensating for a specific functional limitation through a one-person, one-tool interaction model.

Disability scholarship has long argued that this framing is incomplete. Bennett et al. \cite{bennett_interdependence_2018} argue that interdependence, not independence alone, better captures how people with disabilities navigate daily life, and should sit at the centre of assistive technology research. Xiao et al. map recurring patterns in ability-diverse collaboration, showing that collaboration is itself the mechanism through which access gets produced \cite{xiao_systematic_2024}. This collaborative reality is evident across settings: blind runners coordinate with sighted guides through continuous, fine-grained communication that weaves together voice cues and bodily movement \cite{barbareschi_speech_2024}, and creators with sensory impairments split the work of video production across multiple stages with trusted partners, from filming through editing and publishing \cite{xiao_understanding_2025}. For many people with disabilities, collaboration is the ordinary infrastructure behind complex professional and creative work.

Human-agent collaboration research offers relevant foundations for understanding this collaborative reality, with frameworks for trust, coordination, and shared agency \cite{amershi_guidelines_2019,horvitz_principles_1999,grudin_tool_2018,yao_human-human_2026} building on grounding theory \cite{clark_grounding_1991}, workspace awareness \cite{dourish_awareness_1992,gutwin_descriptive_2002}, and knowledge boundary management \cite{carlile_transferring_2004}. Yet these frameworks have not engaged with ability-diverse settings, and carry assumptions that need rethinking when ability differences are present.

In this position paper, we propose a three-layer framework that bridges accessibility and human-agent collaboration. Layer 1, Channelling, concerns modality-adapted, equivalent access to task-relevant information. Layer 2, Coordinating, concerns how AI can mediate workflows, communication, and handoffs among collaborators with different abilities. Layer 3, Co-Creating, moves beyond support to consider AI as a bounded contributor that advances shared goals alongside human partners. Grounded in the Ability-Diverse Collaboration framework \cite{xiao_systematic_2024}, grounding theory \cite{clark_grounding_1991}, and Carlile's 3T framework \cite{carlile_transferring_2004}, together these layers extend both fields by centring interdependence in people with disabilities' everyday practices and by treating triadic collaboration as an explicit design target.

%% file: Chapters/2_Related_Work.tex
\section{Background and Positioning}

Ability-diverse collaboration, where collaborators bring fundamentally different abilities to shared work, differs from typical collaboration in two key ways \cite{xiao_systematic_2024}. First, collaborators may not enjoy equal access to information, nor possess congruent knowledge concerning the content and process of collaboration. Second, this inherent asymmetry in abilities and information access can engender distinct roles and potentially divergent goals. These two features challenge foundational assumptions in the theories most commonly used to explain collaboration.

Clark and Brennan's \cite{clark_grounding_1991} theory of grounding describes how collaborators establish mutual understanding through a collective process shaped by the communication medium's constraints: copresence, visibility, audibility, cotemporality, and others. A central principle is least collaborative effort: participants minimise the total work needed to reach mutual understanding. In ability-diverse collaboration, however, these constraints are not properties of the medium alone but of the person-medium coupling. A video call affords visibility, but not for a blind collaborator. A voice channel affords audibility, but not for a Deaf collaborator. The first feature of ability-diverse collaboration, unequal information access, means that grounding costs become fundamentally asymmetric. The least-effort path for the group may concentrate information gatekeeping in one party, efficient in aggregate but constraining for the person with a disability.

Carlile's \cite{carlile_transferring_2004} 3T framework describes three progressively complex knowledge boundaries: syntactic (a shared lexicon suffices), semantic (interpretive differences require translation), and pragmatic (conflicting interests mean knowledge is ``at stake'' and must be transformed). The second feature of ability-diverse collaboration, that asymmetry creates distinct roles and divergent goals, maps directly onto these pragmatic boundaries. The knowledge at stake is not only domain-specific expertise but ability-specific: the embodied, situated understanding of how to perceive, move, and work. When a collaboration must be renegotiated because abilities or roles change, this knowledge carries real costs, and mismatches between the boundary type and the process used produce costly failures.

Human-agent collaboration research has been exploring agents as collaborative partners, building on grounding theory and workspace awareness \cite{dourish_awareness_1992,gutwin_descriptive_2002}, and developing frameworks for trust, coordination, and shared agency \cite{amershi_guidelines_2019,horvitz_principles_1999,yao_human-human_2026}. This creates a natural opportunity: AI could address both features of ability-diverse collaboration, reducing information asymmetry and mediating the distinct roles that ability differences create. However, current frameworks presume shared perceptual access to the same evidence, a common modality for presenting awareness information, and a dyadic structure. In ability-diverse settings, AI typically enters an existing partnership, creating a triad with fundamentally different coordination demands. Addressing these challenges requires a framework that accounts for both: how AI can reconfigure the grounding conditions so that information access is no longer dependent on a single collaborator, and how AI can enter the negotiation of roles and goals without overriding the agency of the people involved.

%% file: Chapters/3_Background.tex
\section{The Three-Layer Framework}
We propose three layers of agent involvement in ability-diverse collaboration, Channelling, Coordinating, and Co-Creating, each progressively deeper (\autoref{fig:teaser}). The layers are nested: Coordinating depends on Channelling, and Co-Creating depends on both. This nesting parallels Carlile's \cite{carlile_transferring_2004} progression from syntactic to semantic to pragmatic boundaries: each layer requires not just more support, but a qualitatively different kind of shared knowledge. We ground each layer in the ADC technology categories and illustrate with examples across disability types and domains.

\subsection{Layer 1: Channelling}
In ability-diverse collaboration, information asymmetry is often the root barrier. A Deaf collaborator in a design team cannot hear verbal feedback during a critique session. A collaborator using switch-based input cannot scan a dense project dashboard at the pace their team assumes. A neurodivergent team member may receive the same meeting notes as everyone else but find the unstructured format unusable for extracting priorities. This asymmetry creates unequal costs: some collaborators must invest significantly more effort to access the same information, rely more heavily on partners to interpret or relay it, or miss information altogether. Over time, these imbalances can constrain agency and reinforce dependency on others. Research on accessible collaborative writing \cite{das_co11ab_2022} and collaboration awareness in document editing \cite{lee_collabally_2022} has begun to address fragments of this problem, but typically for a single user rather than a collaborative pair.

At this layer, AI functions as an ability supporter, providing equivalent, modality-adapted information to all collaborators. This does not mean identical information, but functionally equivalent access sufficient to support informed decision-making. Real-time captioning can give a Deaf team member access to rapid-fire design discussion. Structured meeting summaries with prioritised actions can give a collaborator with executive function difficulties a usable overview of what was decided. AI-generated descriptions can give a blind collaborator independent access to visual workspace state.

In Clark and Brennan's \cite{clark_grounding_1991} terms, this layer reconfigures the grounding conditions so that the costs of establishing mutual understanding no longer fall disproportionately on one party. The key shift is from delegation to informed direction. When a collaborator has independent access to task-relevant information, the collaboration dynamic changes: rather than asking ``what did I miss?'', they can direct ``let us revisit that decision.'' However, establishing shared information does not resolve how to coordinate around it. For that, the collaboration must move to the next layer.

\subsection{Layer 2: Coordinating}
Even when all collaborators can access the same content, ability-diverse collaboration requires substantial coordination work: negotiating roles, managing handoffs, tracking each other's activities, and resolving differences in judgment. This coordination work is often invisible and burdensome, what Branham and Kane \cite{branham_invisible_2015} call the ``invisible work of accessibility'' in mixed-ability workplaces.

At this layer, AI functions as a communication supporter, mediating workflow between collaborators with different abilities. This means maintaining task state across ability boundaries: providing awareness cues adapted to each collaborator's sensory and cognitive modality, tracking energy-aware task distribution in a team where a member with chronic fatigue works alongside non-disabled colleagues, or supporting structured handoffs when collaborators share a single device, a common setup in low-resource contexts \cite{barbareschi_social_2020,xiao_understanding_2025}. Prior work on paired remote sighted assistance \cite{xie_are_2023} shows that even human-human coordination in these settings is effortful; AI-mediated coordination could reduce this burden across disability types.

This layer addresses workspace awareness \cite{gutwin_descriptive_2002} in ability-diverse settings, where the standard assumption that all collaborators can monitor the same shared workspace does not hold. In Carlile's \cite{carlile_transferring_2004} terms, it develops semantic capacity: not just transmitting information, but creating shared meanings around what differences in pace, process, and ability mean for how the team works together. The shift is from ad hoc verbal negotiation to structured, AI-mediated workflow continuity, reducing the invisible labour that falls disproportionately on people with disabilities and their partners. However, when differences generate not just interpretive challenges but conflicting interests, when changing one's practice carries real costs, a different process is needed.

\subsection{Layer 3: Co-Creating}
In Channelling and Coordinating, the collaboration model remains ability sharing: the primary goal-holder directs, and both human collaborator and AI serve supporting roles. Co-Creating marks a shift toward ability combining, in which AI contributes as a third participant with distinct capabilities directed toward a shared goal.

At this layer, AI generates, suggests, and contributes. In collaborative music-making between musicians with different motor abilities, AI could generate rhythmic accompaniment patterns that complement each performer's physical playing style. In a workplace team where a member with ADHD struggles to synthesise discussions into structured plans, AI could draft proposals that the whole team iterates on, contributing executive function support as a shared capability rather than an individual accommodation. In design prototyping where a collaborator with limited hand mobility works alongside a non-disabled partner, AI-driven fabrication tools could execute fine motor tasks under the disabled collaborator's direction. Emerging work on human-AI collaboration in remote sighted assistance \cite{lee_opportunities_2022,yu_humanai_2024} explores related configurations, though not yet in co-creative contexts.

The critical design requirement is that AI contributions are proposals, not decisions: transparent, modifiable, and clearly distinguished from human-authored work. In Carlile's \cite{carlile_transferring_2004} terms, this layer faces a pragmatic boundary: AI's entry introduces novelty that creates new dependencies and potentially conflicting interests. What is at stake is deeply personal: one's way of working, one's creative authority, one's professional identity. Research with people with disabilities consistently shows that they maintain creative control even in heavily collaborative workflows \cite{xiao_understanding_2025}. Each new triadic configuration requires its own iterative process of negotiating roles and authority, not the assumption that what worked before will transfer. Extending the ADC notion of ``ability provider'' to a non-human agent is a deliberate theoretical move: AI does not share abilities in the same relational, embodied sense that human collaborators do \cite{bennett_care_2020}. But the functional role, contributing a required capability toward a shared goal, maps onto the ability combining model, and naming it as such allows us to reason about triadic configurations that current CSCW frameworks do not address.

%% file: Chapters/4_Method.tex
\section{Research Agenda}
The framework opens questions at the intersection of accessibility and human-agent collaboration. We highlight four that we believe are especially pressing.

\textbf{What counts as equivalent Channelling across different abilities?} A text description of a video frame conveys different things than seeing it. A structured meeting summary conveys different things than the live discussion. How should AI calibrate detail level and format, and who controls that calibration? How do cultural and linguistic contexts shape what counts as adequate access?

\textbf{How does reducing information asymmetry reshape collaboration?} If Channelling gives all collaborators equivalent access, do established roles shift? Does a collaborator who can now independently access previously gatekept information take on new tasks, or do existing divisions of labour persist? What new tensions emerge?

\textbf{How is agency negotiated in triadic configurations?} When does AI shift from supporting collaboration to transforming it? How do collaborators negotiate creative ownership when AI has generated content that both humans refine? This question is especially sharp for people with disabilities who have fought to maintain their creative authority. Research on algorithm aversion \cite{dietvorst_algorithm_2015} and complementary team performance \cite{bansal_does_2021} suggests that trust calibration in human-AI teams is already complex; ability-diverse settings add further dimensions.

\textbf{What extensions to CSCW theory are needed?} The ``remote collaborator'' metaphor is productive, but a collaborator that perceives through multiple modalities simultaneously, never tires, and has no lived experience of disability is a fundamentally different kind of partner. How do common ground and workspace awareness need rethinking for triadic, ability-diverse settings \cite{clark_grounding_1991,gutwin_descriptive_2002}, and what new concepts might be required? How should the notion of knowledge ``at stake'' \cite{carlile_transferring_2004} be extended when what is at stake is not domain expertise but embodied, ability-specific ways of working? Parasuraman et al.'s taxonomy \cite{parasuraman_model_2000} of automation levels may offer a complementary lens for specifying AI's degree of involvement across the three layers.

%% file: Chapters/6_Discussion.tex
\section{Discussion}
The framework makes two claims. First, accessibility should be understood as collaborative infrastructure, not only individual assistive capability. When access is generated through collaboration, as empirical research consistently demonstrates \cite{xiao_systematic_2024,bennett_interdependence_2018,wang_accessibility_2018}, AI systems must be designed to facilitate that collaboration rather than merely to address individual deficiencies. Clark and Brennan's \cite{clark_grounding_1991} grounding theory helps make this concrete: accessibility barriers are, at their core, asymmetries in the costs of establishing mutual understanding. Channelling addresses these asymmetries, but Coordinating and Co-Creating recognise that equitable access alone is not enough; the coordination structures and creative roles built around those asymmetries must also be renegotiated.

Second, the maturity of AI in disability contexts should be evaluated through socio-technical outcomes: role transparency, repairability of breakdowns, and fair distribution of coordination labour, not only individual assistive task performance. Carlile's \cite{carlile_transferring_2004} framework adds a specific warning here: when the process used does not match the type of boundary faced, failures surface downstream. Evaluating AI in ability-diverse collaboration must therefore attend to whether the system supports the right kind of shared knowledge at each layer, not just whether it provides accurate outputs.

These claims carry practical implications. System designers should make collaboration states visible: roles, intentions, uncertainties, handoff statuses, and escalation pathways. Based on current tools that deal with parts of this problem \cite{lee_collabally_2022,das_co11ab_2022,das_simphony_2023}, we argue that future systems should surface collaboration metadata: who contributed what, what the AI is uncertain about, and where human judgement is needed, in ways that are appropriate for each participant's abilities and modalities.

Another implication has to do with governance. If triadic co-agency is a goal for design, organisations need policies and interfaces that make it easy to negotiate and contest AI's role. Without these, advanced AI may replicate paternalistic patterns using a novel lexicon, a concern frequently articulated by disability scholars regarding assistive technologies \cite{hofmann_living_2020,mankoff_disability_2010}.

We note several limitations. The framework is conceptual; the layers need to be tested through design research and user studies with people with disabilities who collaborate with AI systems. The mapping between our three layers and Carlile's three boundaries is productive but not exact: ability-diverse collaboration may present boundary configurations that do not fit neatly into the syntactic-semantic-pragmatic progression, particularly when multiple disability types are present. The framework has not yet addressed how its layers interact with power asymmetries beyond ability differences (e.g., employer-employee, expert-novice). 

%% file: Chapters/7_Conclusion.tex
\section{Conclusion}
Human-agent collaboration research has made real progress on trust, coordination, and shared agency, but has largely overlooked a basic fact: for many people with disabilities, collaboration is not occasional but structural, the ordinary infrastructure through which complex work gets done. Accessibility research understands this deeply but has not yet engaged with AI agents as collaborative partners. The question that connects both fields is deceptively simple: what happens when an AI agent joins a collaboration already shaped by difference, interdependence, and hard-won agency?

Our three-layer framework proposes that AI's involvement is a progression, not a single relationship. Channelling establishes shared informational ground across abilities, reconfiguring the grounding conditions so that mutual understanding no longer depends on a single gatekeeper. Coordinating mediates handoffs and reduces the invisible coordination labour that ability-diverse partnerships absorb, building shared meanings across different ways of working. Co-Creating contributes as a bounded participant whose proposals human collaborators evaluate and refine on their own terms, navigating the pragmatic boundaries where roles, interests, and creative authority must be negotiated.

The framework is conceptual and needs empirical grounding through design research with people with disabilities who collaborate with AI systems, engagement with power asymmetries beyond ability differences, and extension across a wider range of disability experiences and professional domains. But we believe its core move, centring interdependence rather than individual deficit, and treating triadic ability-diverse collaboration as an explicit design target, opens productive ground for both fields to build on.